\documentclass[
reprint,
 amsmath,amssymb,
 aps,
]{revtex4-2}

\usepackage{graphicx}
\usepackage{amsmath}
\usepackage{dcolumn}
\usepackage{bm}
\usepackage [english]{babel}
\usepackage [autostyle, english = american]{csquotes}
\usepackage[dvipsnames]{xcolor}

\begin{document}


\title{Microscopic view of materials properties of liquids:\\An atomic scale perspective}

\author{Jaeyun Moon}
\email{To whom correspondence should be addressed; E-mail: jaeyun.moon@ufl.edu}
 \affiliation{
Department of Mechanical and Aerospace Engineering,\\University of Florida, Gainesville, FL 32611, USA\\
Department of Materials Science and Engineering,\\University of Florida, Gainesville, FL, 32611, USA}

 

\date{\today}
\clearpage

\begin{abstract}
Microscopic understanding of liquid properties is essential for advancing a wide range of applications from energy applications such as nuclear reactors and batteries to biomedical applications including drug delivery and microfluidics. However, intrinsic dynamic disorder and lack of structural periodicity in liquids have presented fundamental challenges in developing rigorous microscopic theories of their thermodynamic and dynamic behavior. Recent breakthroughs in computational power and experimental metrologies have driven significant progress in unraveling the complex atomic scale dynamics of liquids. In this Review, we provide a brief historical context of liquid state physics and explore recent advances through theoretical, computational, and experimental approaches. For theoretical and computational approaches, instantaneous normal mode and velocity autocorrelation function calculations are discussed. For experiments, we focus on X-ray and neutron scattering techniques that probe liquid dynamics at the atomic level. Finally, we highlight emerging opportunities and future directions in the study of liquid atomic dynamics.

\end{abstract}


\maketitle
\tableofcontents

\section{Introduction}

Materials properties of liquids are critical in a wide range of applications, from energy technologies such as thermal storage and nuclear reactors \cite{li_sensible_2016, williams_assessment_2006} to pharmaceutical and drug-delivery systems \cite{adawiyah_ionic_2016}. For example, due to explosion in our need for electricity to power data centers in recent years, next generation designs of nuclear reactors have recently garnered immense interests from both government and industry sectors. Here,  thermophysical properties of liquid coolants impact the rate at which heat is extracted from the core, directly affecting the overall efficiency and performance of electricity generation and storage \cite{roper_molten_2022, magnusson_review_2020, sabharwall_advanced_2014}. Despite their importance in numerous applications such as nuclear reactors, our microscopic understanding of liquid materials properties remains far less developed than that of solids and gases, preventing disruptive technological advances utilizing liquids \cite{allen_computer_1987, wallace_statistical_2002, stillinger_energy_2015}.

In solids (particularly crystalline solids), atoms vibrate about well-defined equilibrium positions. Fixed equilibrium positions enable perturbative theoretical approaches (e.g., Taylor expansion), which have been highly successful in describing atomic dynamics and the resulting materials properties. For example, phonon quasi-particles (quantized lattice vibrations) provide a powerful framework for predicting heat capacity, entropy, thermal expansion, and thermal conductivity in dielectric solids, often with first-principles accuracy \cite{togo_first_2015, lindsay_survey_2018}.

At the other extreme, dilute gases are characterized by weak interatomic interactions, and their dynamics are governed primarily by atomic or intermolecular collisions and internal degrees of freedom. This provides the basis for kinetic theory \cite{bernoulli_hydrodynamica_1738}, which successfully describes transport properties such as viscosity and thermal conductivity, as well as thermodynamic quantities such as temperature and pressure.

Liquids, however, lie between these two regimes and lack the simplifying features that make solids and gases tractable. Their structures are dynamically disordered and lack spatial periodicity without a well-defined structure unlike solids. In contrast to dilute gases where atoms are nearly free, atoms in liquids remain strongly interacting due to high density despite the absence of long range order. Consequently, the complex atomic dynamics in liquids has led to difficulties in rigorous theoretical development \cite{trachenko_purcell_2020, brazhkin_collective_2014, debenedetti_supercooled_2001, tanaka_revealing_2019, moon_heat_2024}.

Reflecting these challenges, Lev D. Landau, together with Evgeny M. Lifshitz, wrote that “Liquids do not allow a calculation in a general form of the thermodynamic quantities or even of their dependence on temperature” \cite{landau_statistical_1958}, suggesting that a rigorous microscopic theory of liquid properties may be fundamentally elusive. Even today, this observation remains largely accurate in characterizing the current state of our understanding of liquid materials properties.

Despite these challenges, there is strong reason for optimism. Advances in high-performance computing and experimental capabilities have created unprecedented opportunities to revisit fundamental questions in liquid-state physics and develop rigorous theories capable of predicting materials properties with quantum mechanical accuracy.

On the computational side, large-scale molecular dynamics simulations can now reach system sizes of billions of atoms and simulation times approaching tens of milliseconds \cite{guo_extending_2022, johansson_micron-scale_2022, scalliet_thirty_2022}. Machine learned potentials have enabled description of interatomic interactions at the level of density functional theories \cite{behler_perspective_2016, deringer_machine_2017, fan_neuroevolution_2021, sosso_understanding_2018, moon_crystal-like_2025, babaei_machine-learning-based_2019}. Materials discovery through machine learning has also enabled exploration of vast chemical configuration space that would otherwise have been nearly impossible \cite{luo_predicting_2023, butler_machine_2018}. These advances enable increasingly direct comparisons between simulations and experiments conducted in conventional laboratory settings \cite{walsh_nanoindentation_2003,jia_pushing_2020,scalliet_thirty_2022}.

Complementing these computational breakthroughs, experimental techniques have also advanced dramatically. Large-scale user facilities such as the Spallation Neutron Source at Oak Ridge National Laboratory and the Linac Coherent Light Source at SLAC National Accelerator Laboratory provide access to probes with angstrom-level spatial resolution and femtosecond temporal resolution \cite{trigo_fourier-transform_2013, monti_ultrafast_2026, shin_ultrafast_2023}. For instance, utilizing a femtosecond X-ray probe, laser-induced melting and disintegration kinetics of gold nanoparticles have been studied, revealing both the nanoscale electron density distribution and lattice structures during melting \cite{shin_ultrafast_2023}. Together, these computational and experimental tools offer an unprecedented platform for testing new ideas and hypotheses. If one has an idea or hypothesis, these tools are readily available to help test it, and could eventually help its crystallization into a rigorous theory without a lot of defects.

In this Review, we present a concise overview of progress in the microscopic understanding of the material properties of liquids, drawing on theoretical, computational, and experimental advances. Unlike many well-established research areas where a small number of standard frameworks may dominate, the study of liquids has produced a wide range of analytical approaches and conceptual perspectives. Our aim is therefore not to provide an exhaustive account, but rather to offer a guide map that helps readers navigate the extensive literature devoted to liquid state physics regarding atomic motion.

We begin our Review with a brief historical perspective on efforts to understand the liquid phase, providing context for the current state of the field. We then discuss modern computational and theoretical approaches utilizing normal mode decomposition and velocity autocorrelation functions for describing atomic dynamics in liquids and their connection to macroscopic material properties. We particularly focus on these as they directly characterize the atomic degrees of freedom in liquids in contrast to approaches that make explicit assumptions on the atomic motion. Next, we review insights gained from X-ray and neutron scattering techniques for probing liquid dynamics at the atomic scale. Finally, we conclude with a general outlook on liquid state physics at the atomic scale.

\section{Brief Historical Account}
\subsection{Birth of kinetic theory of gas}
Our physical and in some respects, philosophical understanding of matter has a long and rich history. The idea that matter is composed of small, indivisible particles, termed atomos, dates back to ancient Greece. Despite the persistence of this conceptual framework for millennia, a rigorous modern atomic theory did not emerge until the early nineteenth century. This breakthrough was articulated by Dalton through what is now known as the law of multiple proportions \cite{dalton_new_1808}. In doing so, he shifted the focus of chemical atomism away from earlier speculative ideas about atomic shape and interatomic forces toward relative atomic weights and quantitative composition, emphasizing the characterization of individual chemical species rather than the classification and rationalization of chemical reactions \cite{giunta_four_2010}.

\begin{figure}
	\centering
	\includegraphics[width=0.49\linewidth]{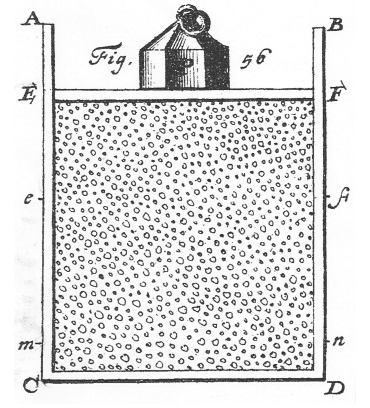}
	\caption{Visualization of gas pressure via a kinetic model of atoms and molecules by Bernoulli in 1738 \cite{bernoulli_hydrodynamica_1738}.}
	\label{fig:Bernoulli}
\end{figure}

One of the earliest natural phenomena explained through the motion of atoms was the behavior of gases. The conceptual basis for the kinetic theory of gases appeared as early as the seventeenth century in various writings \cite{boyle_defence_1662}. In the early eighteenth century, several semi-kinetic models were proposed by Euler and others, although these were based on the notion that atoms in a gas had vibrational or rotational motion and were suspended in ether \cite{brush_kind_1976}. The first quantitative and recognizably modern kinetic description emerged with Bernoulli in 1738, when he proposed that gas pressure is proportional to the square of molecular velocity (see Fig. \ref{fig:Bernoulli}) \cite{bernoulli_hydrodynamica_1738}.

However, the kinetic theory could not fully develop while heat was still widely interpreted as a fluid-like substance known as caloric. Only after heat came to be understood as a manifestation of atomic motion did the theory gain traction. As a result, major developments by Krönig, Clausius, Joule, Maxwell, and Boltzmann occurred primarily in the second half of the nineteenth century, many decades after Dalton’s formulation of atomic theory. Maxwell-Boltzmann distribution of atomic velocities of gases became among the first statistical and probabilistic approach to studying physical phenomena. As a result of better understanding of atoms and developmentof kinetic theory of gases, phases of matter were no longer viewed as static equilibria between attractive interatomic forces and repulsive interactions involving caloric, but instead as consequences of the dynamic balance between attractive and repulsive interatomic forces \cite{giunta_four_2010}. These foundational ideas involving atomic motion in gases profoundly reshaped our modern understanding of matter and subsequently, liquids. With these general historical discussions, we now provide a more specific context around liquids, the central theme of our Review. 

\subsection{Liquids from gas perspective}

In parallel with efforts to understand atomic motion in gases and to develop the kinetic theory of gases, substantial work was also devoted to understanding the nature and classification of the phases of matter. During the seventeenth and eighteenth centuries, interest in liquids and gases at high temperatures and pressures grew rapidly, in part driven by technological developments such as the steam engine. One particularly remarkable contribution was Cagniard de la Tour’s discovery of critical phenomena.

In the 1820s, Cagniard de la Tour, motivated by interests in acoustics, conducted experiments in which a flint ball was placed inside a partially filled sealed digester containing a liquid. As the vessel rolled, a splashing sound occurred when the ball crossed the interface between the liquid and its vapor. Upon heating the system well beyond the liquid’s boiling point, however, Cagniard de la Tour observed that this splashing sound disappeared above a specific temperature \cite{de_la_tour_expose_1822}. This observation marked the first experimental identification of what is now known as the supercritical fluid state.

The discovery generated significant debate about how to define and categorize liquids, gases, and what some referred to as the “Cagniard de la Tour state.” Prominent scientists including Herschel, Faraday, Whewell, Dumas, and Mendeleev discussed the implications of these observations \cite{herschel_preliminary_1830, faraday_xvii_1823, faraday_selected_1971, mendelejeff_ueber_1861}. Herschel argued that Cagniard de la Tour’s work suggested the absence of a sharp boundary between the three classical states of matter, writing that the solid, liquid, and gaseous states likely represent stages in a gradual progression rather than sharply separated categories.

Faraday and Whewell later corresponded about how best to describe and name this newly observed state. Whewell, well known for coining terms such as scientist and physicist, proposed calling these supercritical states vaporiscents and described liquids crossing the critical point as becoming disliquified, implying that the liquid state was being destroyed. Faraday found these suggestions unsatisfactory. Mendeleev offered a different interpretation, suggesting instead that liquids transform into vapors. Only a few years later, Mendeleev would publish the work that ultimately led to the modern periodic table \cite{mendelejew_uber_1869}.

\begin{figure}
	\centering
	\includegraphics[width=0.8\linewidth]{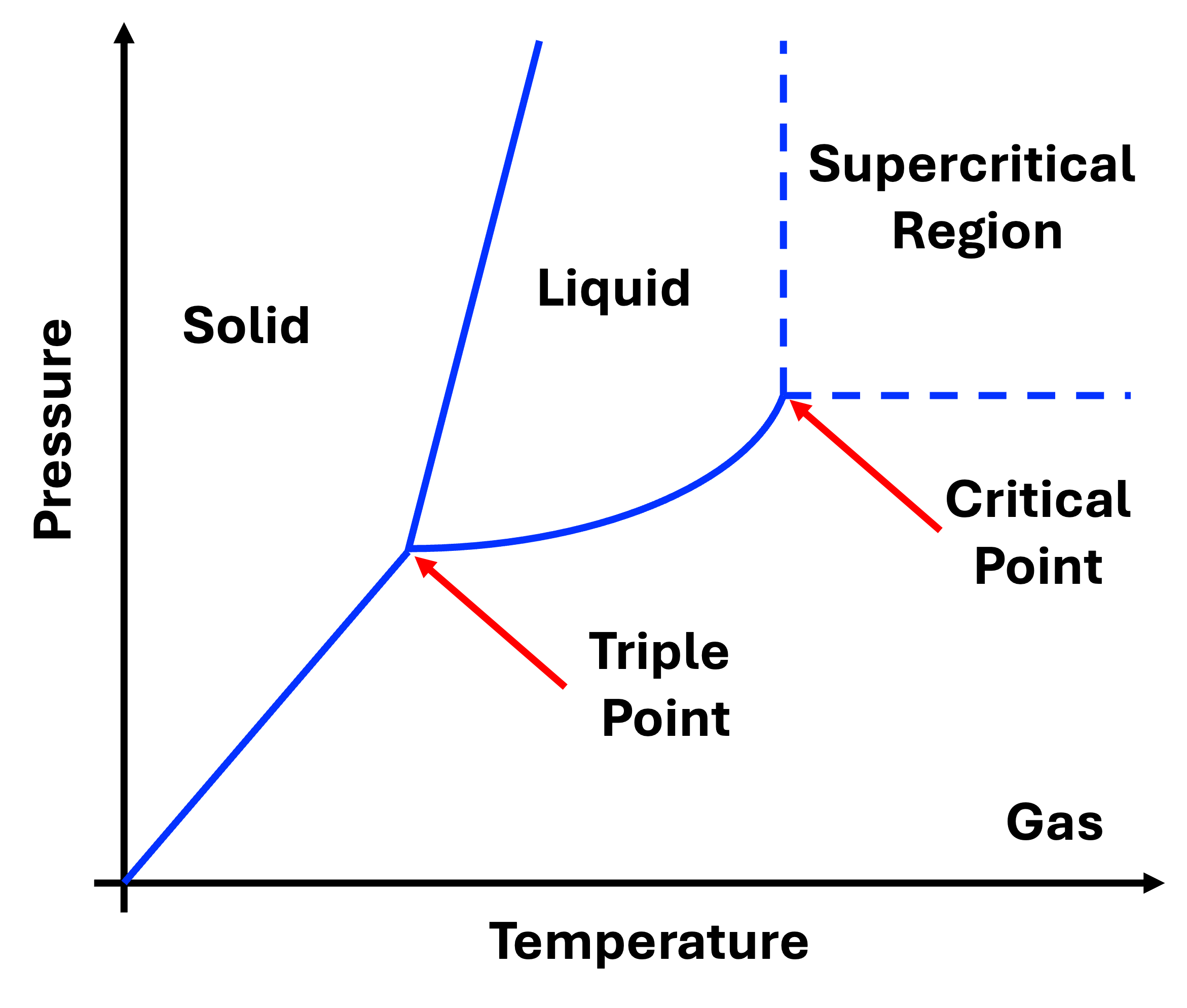}
	\caption{A representative phase diagram (temperature vs. pressure) of matter. Slopes of the boundaries depend on the materials.}
	\label{fig:Phase}
\end{figure}

Further clarification came from Andrews’ liquefaction experiments on carbon dioxide in the 1860s. Andrews demonstrated that liquids and gases approach a common fluid state at a specific temperature and pressure, which he termed the critical point. He presented these findings in his celebrated Bakerian Lecture to the Royal Society titled, “On the continuity of the gaseous and liquid states of matter” \cite{andrews_bakerian_1869, tait_scientific_1889}. These studies established the modern concept of the critical point and supercritical state and helped shape the phase diagrams used today (Fig. \ref{fig:Phase}).

In 1873, van der Waals showed in his doctoral dissertation that Andrews’ experimental observations could be explained by a simple extension of the ideal gas law that incorporated molecular attraction and repulsion \cite{van_der_waals_over_1873}. Notably, van der Waals’ thesis also bore the title “On the continuity of the gaseous and liquid states”. During this period, the term fluid increasingly came to encompass both gases and liquids, as reflected in later works such as Ref. \cite{onnes_theorie_1881}, and the van der Waals equation eventually became known as an “equation of fluids.” Beyond its immediate impact on thermodynamics, van der Waals’ work also contributed to the development of the hard-sphere paradigm that remains widely used in studies of soft matter, granular materials, gases, and liquids \cite{chandler_van_1983, dyre_simple_2016}.

It is also noteworthy that the Navier–Stokes equations describing the flow of viscous fluids, applicable to both liquids and gases, are developed during roughly the same period as the pioneering studies of Cagniard de la Tour, Andrews, and van der Waals on supercritical phenomena.

\subsection{Liquids from solid perspectives}

In addition to considering liquids from gas perspectives as fluids, liquids have also been considered from solid perspectives as condensed matters. It is not clear when similarity between solid and liquid states at low temperatures became the basis for theoretical developments for liquids. Capillary action first recorded by da Vinci in the 15th century is a demonstration of the importance of strong intermolecular interactions in liquids, similar to atoms in solids but unlike gases \cite{da_vinci_manuscripts_1490}. In addition to the critical phenomena, van der Waals also investigated microscopic mechanisms of capillary action in his thesis \cite{van_der_waals_over_1873}.

More systematic theoretical treatments that considered solids and liquids together began as early as 1903, when Mie modeled liquids as an array of harmonic oscillators located on lattice sites \cite{mie_zur_1903}. For nearly a century thereafter, many researchers modeled liquids as systems in which atoms occupy mean positions similar to those in crystals, or as crystalline structures containing interstitial defects \cite{mott_resistance_1934,lennard-jones_critical_1937,granato_specific_2002}. These are largely regarded as lattice theories of the liquid phase \cite{rowlinson_lattice_1951}. Within this picture, various thermodynamic properties of liquids have been investigated.

During the same period of the lattice theory developments, Frenkel proposed a different viewpoint in which solids and liquids are treated as a continuity \cite{frenkel_uber_1926,frenkel_continuity_1935,frenkel_liquid_1937,frenkel_kinetic_1947}. He grouped these phases together under the term “condensed bodies” though the terms including condensed matter and condensed bodies have existed previously. Notably, one of his papers is titled “Continuity of the Solid and the Liquid States”, highlighting a conceptual contrast with the earlier works of Andrews and van der Waals. In his work, Frenkel proposed that a material behaves like a solid when the characteristic observation time is shorter than $\tau = \eta/\mathcal{N}$ and like a liquid when it is longer. Here, $\eta$ is the viscosity and 
$\mathcal{N}$ is rigidity modulus \cite{frenkel_liquid_1937}. Frenkel's seminal ideas have led to many recent works considering thermodynamics of liquids from phonon quasi-particles as in solids \cite{trachenko_heat_2011, bolmatov_phonon_2012,andritsos_heat_2013, trachenko_collective_2016, wang_direct_2017, yang_emergence_2017, brazhkin_liquid-like_2018, tomiyoshi_heat_2019, khusnutdinoff_collective_2020, Kryuchkov_universal_2020, zaccone_universal_2021, baggioli_explaining_2021}.

\begin{figure}
	\centering
	\includegraphics[width=1\linewidth]{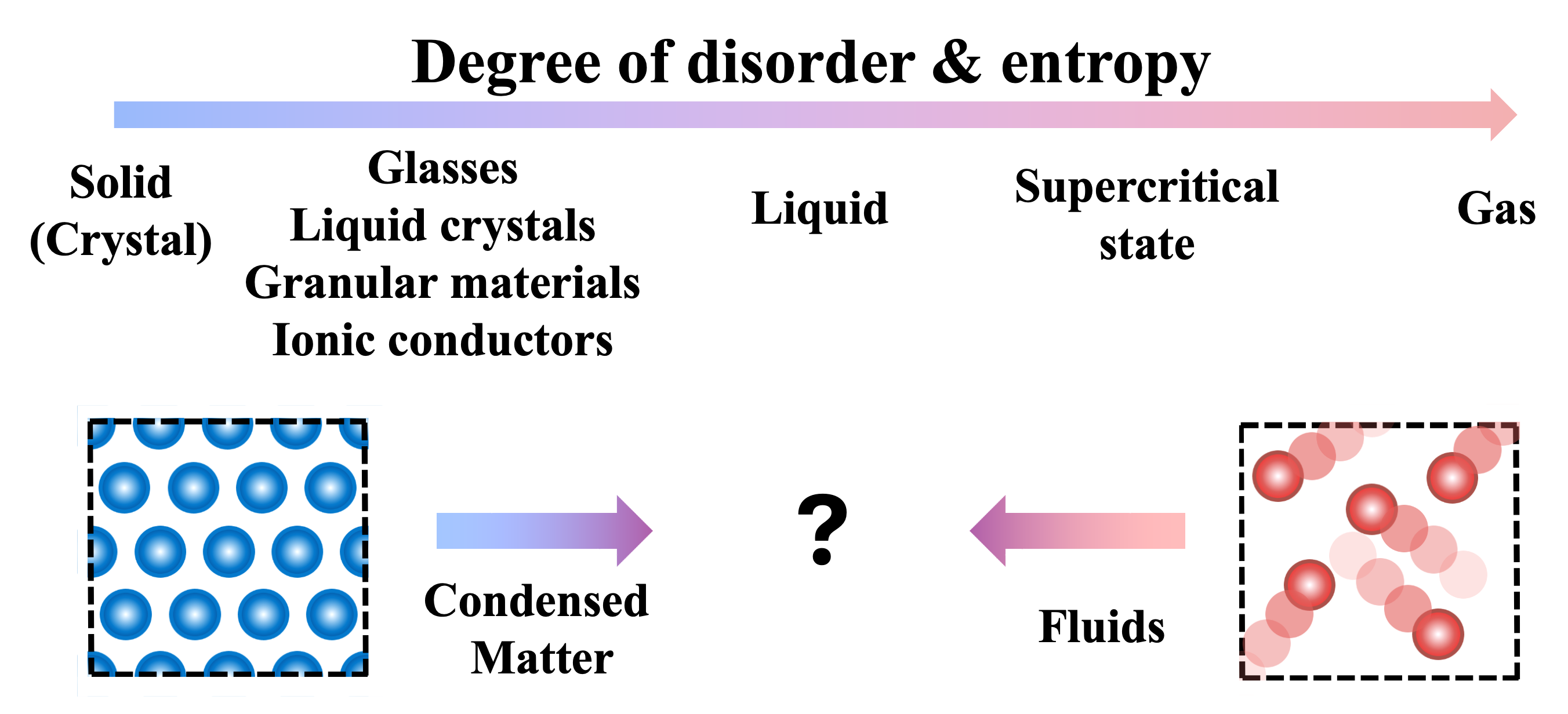}
	\caption{A schematic demonstrating spectrum of matter from solid to gas and how liquids are often viewed: either a condensed matter or a fluid.}
	\label{fig:Spectrum}
\end{figure}

Whether regarded as fluids or as condensed matter, liquids have historically been studied from either a gas-like or a solid-like perspective—two viewpoints that are often conceptually incompatible (see Fig. \ref{fig:Spectrum}). As discussed in more detail below, many modern studies, particularly theoretical and computational ones, still follow these historical frameworks. In parallel, liquids are commonly taught within separate disciplinary contexts—either as fluids in courses such as fluid mechanics or as condensed matter in courses on condensed matter physics. In our view, the very success of these two paradigms has also created a bottleneck in developing a broader microscopic understanding of atomic dynamics in complex matter, including liquids, by shaping education, constraining conceptual frameworks, and limiting the range of analytical approaches that are typically considered.

With this historical background in mind, we next discuss modern computational, theoretical, and experimental approaches for studying liquids at the atomic scale. We begin by introducing the normal mode formalism, traditionally used to describe phonons in solids, which has recently been extended to other phases of matter and to the analysis of velocity autocorrelation functions.





\section{Normal Modes}

In simple, perfect solids, atoms vibrate around their equilibrium positions periodically. Normal mode (sinusoidal planewave) decomposition of atomic motion in solids provides a natural microscopic framework for understanding atomic dynamics in solids. In the literature, particularly in undergraduate/graduate solid-state physics textbooks, these normal modes are commonly used interchangeably with the concept of phonons in crystalline materials \cite{ashcroft_solid_1976}. \\

Over the past few decades, significant efforts have been made to extend the normal mode formalism beyond crystalline solids to liquids and even dilute gases, with varying degrees of success in describing their atomic dynamics. In systems that lack well-defined equilibrium positions, instantaneous atomic configurations are instead used as the reference for normal mode calculations. To distinguish these approaches, the literature commonly refers to equilibrium normal modes, based on equilibrium atomic positions, and non-equilibrium or instantaneous normal modes, which are constructed from instantaneous configurations. In the following, we discuss the equilibrium normal mode formalism in greater detail and current progress in the instantaneous normal mode framework. 

\subsection{Equilibrium Normal Modes}
Liquids typically exist well above Debye temperatures such that atomic dynamics can be considered as classical in nature with some exceptions including liquid Helium at cryogenic temperatures. For direct comparisons between equilibrium normal modes and instantaneous normal modes (for liquids), we discuss here a classical treatment of equilibrium normal mode formalism \cite{dove_introduction_1993}. Consider a structure with an arbitrary number ($n$) of atoms per unit cell with different atomic masses and in three dimensions. There are $3n$ branches per wavevector, of which three are acoustic branches (longitudinal and two transverse) and $3n-3$ are optical branches.

The equation of motion for atom $j$ in the unit cell $p$ under harmonic approximation is given by
\begin{equation}
    m_j\ddot{\boldsymbol{u}}_{j,p}(t)=-\sum_{j',p'}\boldsymbol{\Phi}_{j,p;j',p'} \cdot \boldsymbol{u}_{j',p'}(t)
    \label{Eq:EOM}
\end{equation}
where matrix elements of the force constants, $\boldsymbol{\Phi}_{j,p;j',p'}$, is given by
\begin{equation}
    \Phi_{j,p;j',p'}^{\alpha \beta} = \frac{\partial^2 U}{\partial u^{\alpha}_{j,p} \partial u^{\beta}_{j',p'}}
    \label{Eq:diff_PE}
\end{equation}
with $\alpha$ and $\beta$ being the Cartesian directions and $U$ being the potential energy. Eq. \ref{Eq:EOM} is essentially a more general description of Hooke's law. 

We assume here a planewave form of the solution as
\begin{equation}
    \boldsymbol{u}_{j,p}(t) = \sum_{\boldsymbol{k},\nu} \boldsymbol{A}(j,\boldsymbol{k}, \nu)e^{i[\boldsymbol{k}\cdot \boldsymbol{r}_{j,p}-\omega(\boldsymbol{k},\nu)t]}
    \label{Eq: PWS}
\end{equation}
where $\boldsymbol{r}_{j,p}$ is the equilibrium position of the atom $j$ in unit cell $p$ and $\boldsymbol{A}(j,\boldsymbol{k}, \nu)$ is the amplitude vector for atom $j$ for mode with wavevector $\boldsymbol{k}$ and branch $\nu$ and is independent of $p$ as the differences in atomic motions between unit cells are described in the exponential phase factor. Plugging in Eq. \ref{Eq: PWS} into Eq. \ref{Eq:EOM} and simplifying, we obtain the following eigenvalue problem
\begin{equation}
    \omega^2(\boldsymbol{k}, \nu) \boldsymbol{e}(\boldsymbol{k}, \nu) = \boldsymbol{D}(\boldsymbol{k}) \cdot \boldsymbol{e}(\boldsymbol{k}, \nu)
    \label{Eq: eigenvalue problem}
\end{equation}
with the eigenvector $\boldsymbol{e}(\boldsymbol{k}, \nu)$ having $3n$ by 1 matrix elements. $\boldsymbol{e}(\boldsymbol{k}, \nu)$ is related to $\boldsymbol{A}(j,\boldsymbol{k}, \nu)$ by
\begin{equation}
    \boldsymbol{e}(\boldsymbol{k}, \nu) = 
    \begin{bmatrix}
        \sqrt{m_1}A_x(1,\boldsymbol{k}, \nu)\\
        \sqrt{m_1}A_y(1,\boldsymbol{k}, \nu)\\
        \sqrt{m_1}A_z(1,\boldsymbol{k}, \nu)\\
        \sqrt{m_2}A_x(2,\boldsymbol{k}, \nu)\\
        \sqrt{m_2}A_y(2,\boldsymbol{k}, \nu)\\
        \sqrt{m_2}A_z(2,\boldsymbol{k}, \nu)\\
        .\\
        .\\
        .\\
        \sqrt{m_n}A_z(n,\boldsymbol{k}, \nu)
    \end{bmatrix}
\end{equation}
$\boldsymbol{D}(\boldsymbol{k})$ is known as the dynamical matrix and is defined as
\begin{equation}
    D^{\alpha \beta}_{jj'}(\boldsymbol{k})=\frac{1}{\sqrt{m_jm_{j'}}}\sum_{p'} \Phi_{j,p;j',p'}^{\alpha \beta}e^{i\boldsymbol{k}\cdot [\boldsymbol{r}_{j'p'}-\boldsymbol{r}_{jp}]}.
    \label{Eq: DM}
\end{equation}

Dynamical matrices can be interpreted as the mass-reduced Fourier transform of the force constant matrices $\Phi_{j,p;j',p'}^{\alpha \beta}$. For each $\boldsymbol{k}$ as shown in Eq. \ref{Eq: DM}, the dynamical matrix is a $3n$ by $3n$ matrix. The eigenvalue matrix $\Omega(\boldsymbol{k},\nu)=\omega^2(\boldsymbol{k},\nu)$ is also a $3n$ by $3n$ matrix with only diagonal elements: $\omega^2(\boldsymbol{k},1), \omega^2(\boldsymbol{k},1), \; ... \;, \omega^2(\boldsymbol{k},3n)$. As evident from Eq. \ref{Eq:diff_PE}, normal modes can be thought of as then a way to probe potential energy landscape (PEL) of the system in the reciprocal space as eigenvalues reflect the curvature of the PEL. 

\begin{figure}
	\centering
	\includegraphics[width=1\linewidth]{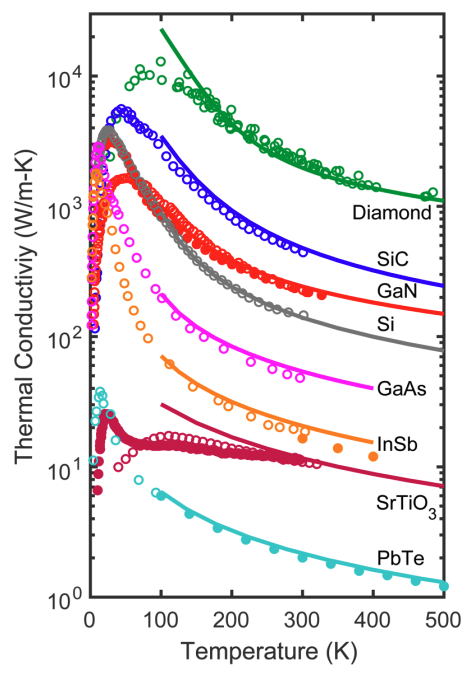}
	\caption{Temperature dependent thermal conductivity for various crystals from diamond to lead telluride. Circles are measurements and curves are first principles calculations based on normal modes. Figure is from Ref. \cite{mcgaughey_phonon_2019}.}
	\label{fig:McGaughey}
\end{figure}

In solids, $\omega(\boldsymbol{k},\nu)$ are physically understood as 
vibrational frequencies and the relationship between $\omega(\boldsymbol{k},\nu)$ and $\boldsymbol{k}$ is known as dispersion relations. Population distribution of the $\omega(\boldsymbol{k},\nu)$ is regarded as the density of states. Knowledge of $\omega(\boldsymbol{k},\nu)$ from solving Eq. \ref{Eq: eigenvalue problem} is a starting point to understanding various thermodynamic and thermal properties of solids including sound velocities, heat capacity, entropy, and thermal conductivity. The accuracy of these property predictions depends on how accurate force constants are.

An example of a materials property prediction using normal modes is shown in Fig. \ref{fig:McGaughey} for thermal conductivity for various crystals. A decent agreement is demonstrated between measurements and calculations based on normal modes from first-principles. Some discrepancies shown may be due to imperfection of the crystals in the measurements. As the focus of this Review regards liquids, for more details of characterizing materials properties of solids through normal modes, Refs. \cite{mcgaughey_phonon_2019, togo_first_2015} are suggested.


Primitive unit cells (the smallest unit cells that describe the lattice) are conventionally used in Eq. \ref{Eq: eigenvalue problem} for crystalline materials, although the use of other unit cells, such as conventional unit cells, is equally valid. The choice of unit cell affects the resulting dispersion relations, which differ from those obtained using the primitive unit cell; however, the density of states remains invariant regardless of the unit cell chosen.

For systems without spatial periodicity, such as amorphous solids, the entire structure under consideration is treated as a single unit cell. Consequently, only the zero wavevector ($\boldsymbol{k}=\boldsymbol{0}$), or $\Gamma$-point, dynamical matrix is used \cite{allen_diffusons_1999,allen_thermal_1989,moon_examining_2021,he_heat_2011,lv_direct_2016,moon_propagating_2018,moon_thermal_2019,agne_minimum_2018,zhou_contribution_2017,larkin_thermal_2014, moon_sub-amorphous_2016,deangelis_thermal_2018}. In this case, the conventional distinction between acoustic and optical branches no longer applies, since no other wavevectors are defined. The phase factor in Eq. \ref{Eq: DM} also vanishes.


\subsection{Instantaneous Normal Modes}

Due to the immense success of utilizing normal mode decomposition of atomic motion in solids for understanding materials properties \cite{togo_first_2015, lindsay_survey_2018, mcgaughey_phonon_2019}, normal modes are often inherently considered as phonons in the literature. However, we note here that the normal mode formalism is at first a mathematical framework. Normal modes being phonons or vibrations is rather a physical interpretation of this mathematical framework applied to solids. Ultimately, the normal mode formalism is a mathematical tool to understand the atomic motion using planewave decompositions. 

Recently, there has been significant interest in extending the normal mode formalism to non-solid systems, where instantaneous snapshot structures are used in place of equilibrium lattices \cite{keyes_instantaneous_1997, stratt_instantaneous_1995, cho_instantaneous_1994, seeley_isobaric_1991, seeley_normalmode_1989, gezelter_can_1997, schirmacher_modeling_2022, zaccone_universal_2021, moon_atomic_2023, moon_normal_2024, la_nave_instantaneous_2000, jin_revisiting_2025, zhang_what_2019, bembenek_instantaneous_1995, sastry_spectral_2001, baggioli_explaining_2021, keyes_unstable_2025}. Normal-mode decomposition of atomic motion in liquids could still be useful because the number of normal modes remains equal to the number of atomic degrees of freedom regardless of the phase of the system. Describing atomic motion in reciprocal space through normal modes may therefore provide useful insight into the underlying degrees of freedom governing dynamics in real space. As in amorphous solids, only the $\Gamma$ point is typically considered. 

\begin{figure}
	\centering
	\includegraphics[width=1\linewidth]{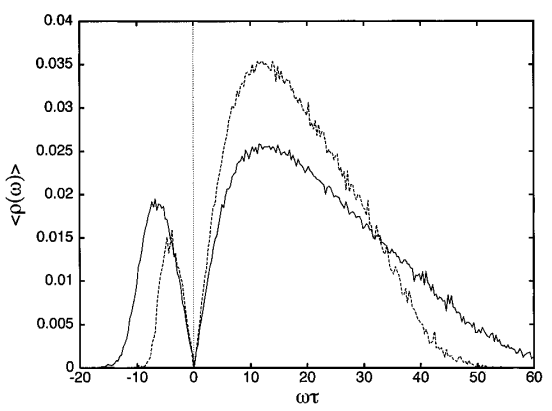}
	\caption{Instantaneous normal mode density of states for Lennard-Jones liquids at two temperatures. Solid curve corresponds to temperature 2.5 times higher than that of dashed curve. Figure is from Ref. \cite{keyes_instantaneous_1997} and $x$-axis is in normalized frequency units. Modes with imaginary frequencies are plotted in the negative $x$-axis. As temperature increases, more modes become imaginary.}
	\label{fig:Keyes}
\end{figure}

As we are using instantaneous snapshots of atomic positions that are not in equilibrium, first expectation is that modes with imaginary frequencies will begin to appear. An example of this is demonstrated in Fig. \ref{fig:Keyes} for the density of states of Lennard-Jones liquids \cite{keyes_instantaneous_1997}. Imaginary modes are plotted in the negative $x$-axis. With increase in temperature, we observe that more modes become increasingly imaginary. Various questions then arise: \textit{What is the physical nature of (instantaneous) normal modes? Especially in liquids? How are real and imaginary modes different from one another? How do they relate to materials properties?} Below, we show some progress in answering these questions. 

In literature, our understanding of the physical nature of equilibrium normal modes in solids dominantly dictates our interpretation of real modes in liquids: These have been considered as equivalent to harmonic oscillators similar to equilibrium normal modes in solids \cite{melzer_instantaneous_2012, stassen_instantaneous_1994}. Therefore, a majority of efforts have been devoted to characterizing the nature of imaginary, unstable modes and how these are related to materials properties, especially transport properties such as diffusion coefficient and viscosity. 

Several models have been proposed to describe self-diffusion coefficients using instantaneous normal modes, motivated by the framework introduced by \citeauthor{zwanzig_relation_1983} \cite{zwanzig_relation_1983}. In this work, Zwanzig attempted to derive the Stokes–Einstein relation starting from the velocity autocorrelation function (VACF) expression for the self-diffusion coefficient, $D = \frac{1}{3} \int \langle \boldsymbol{v}(t) \cdot \boldsymbol{v}(0) \rangle dt$. 

In this picture, atomic motion in liquids is described as oscillatory motion within basins of the potential energy landscape, where the system remains trapped until it encounters a saddle point or bottleneck and transitions to a neighboring basin. The time dependence of the VACF is therefore approximated as a superposition of normal-mode contributions, each proportional to $\cos(\omega t)$ while the system resides within a basin. Transitions between basins interrupt this motion, and different modes are assumed to persist for different durations. This effect is incorporated through an exponential factor $e^{-t/\tau_\omega}$ that represents the waiting-time distribution associated with each mode.

With these assumptions, the self-diffusion coefficient can be written as
\begin{equation}
\begin{split}
    D &= \frac{k_BT}{3mN} \int_0^\infty dt \sum_\omega \cos{(\omega t)} e^{-\frac{t}{\tau_\omega}} \\
    &= \frac{k_BT}{3mN} \sum_\omega \frac{\tau_\omega}{1+\omega^2\tau_\omega^2}
    \label{Eq: Zwanzig_D}
    \end{split}
\end{equation}
where the summation runs over all $3N$ normal modes. Evaluating Eq. \ref{Eq: Zwanzig_D} therefore requires knowledge of both the normal-mode frequencies $\omega$ and the corresponding lifetimes $\tau_\omega$. In Zwanzig’s original work, the discrete sum was replaced by a modified Debye model, and the relaxation times $\tau_\omega$ were estimated from the longitudinal and shear viscosities. However, the applicability of the Debye model to liquids is questionable, since liquids no longer behave as purely elastic media.

Keyes and others proposed an alternative interpretation of Zwanzig’s diffusion model by connecting hopping events between basins of the potential energy landscape to the imaginary-frequency component of the instantaneous normal mode (INM) spectrum \cite{seeley_isobaric_1991, keyes_unstable_1994}. In this framework, the self-diffusion coefficient is suggested to take the form
\begin{equation}
    D=c\langle \omega \rangle_{I} f_I
\label{Eq: Seeley}
\end{equation}
where $c$ is a constant, $\langle \omega \rangle_I$ is the average imaginary-mode frequency, and $f_I$ is the fraction of imaginary modes among all modes. Decent agreement between predictions from Eq. \ref{Eq: Seeley} and independent molecular dynamics calculations was demonstrated (not shown here) \cite{seeley_isobaric_1991, keyes_unstable_1994}. However, this interpretation has limitations. At high temperatures, even crystalline solids without atomic diffusion can exhibit significant displacements from equilibrium lattice sites, which lead to the appearance of imaginary modes. This results in discrepancies with Eq. \ref{Eq: Seeley} \cite{gezelter_can_1997, bembenek_instantaneous_1995, clapa_localization_2012}.

An alternative perspective suggests that only delocalized modes contribute meaningfully to long-range diffusion. Localized modes primarily induce local rearrangements and therefore contribute negligibly to net atomic transport, whereas delocalized modes can facilitate large-scale structural rearrangements that lead to diffusion \cite{clapa_localization_2012}. Accordingly, diffusion is proposed to depend on the delocalized fraction of imaginary modes,
$f_{I,DL} = f_I - f_{I,L}$, where $f_{I,L}$ denotes the fraction of localized imaginary modes. In this model, the diffusion coefficient follows
\begin{equation}
D \sim e^{\frac{A}{T[a\ln(f_{I,DL}) + b]}}
\label{Eq: Clapa}
\end{equation}
where $A$ is an activation energy, $T$ is temperature, and $a$ and $b$ are constants.

\begin{figure}
	\centering
	\includegraphics[width=1\linewidth]{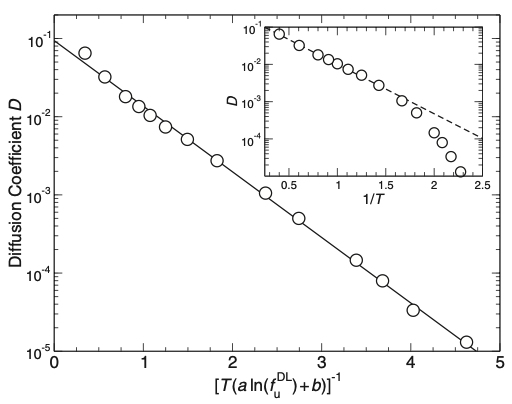}
	\caption{Temperature dependent diffusion coefficient (circles) for Kob-Andersen liquids. Solid line is the fitted, proposed relation based on Eq. \ref{Eq: Clapa}. The inset shows that the data spans both Arrhenius and non-Arrhenius temperature dependence regions. The figure is from Ref. \cite{clapa_localization_2012}.}
	\label{fig:Starr_D}
\end{figure}

Comparisons between Eq.~\ref{Eq: Clapa} and diffusion coefficients obtained from molecular dynamics simulations have been reported for the binary Kob–Andersen (KA) glass-forming liquid \cite{kob_testing_1995}, as shown in Fig.~\ref{fig:Starr_D} \cite{clapa_localization_2012}. The proposed expression captures the diffusion behavior over a broad temperature range spanning this crossover from Arrhenius (i.e., independent atomic motion) to super-Arrhenius dynamics (i.e., correlated atomic motion). Although these results suggest promising connections between instantaneous normal modes and atomic diffusion, further theoretical developments are required to establish a more rigorous framework.

In developing Eq. \ref{Eq: Clapa}, it was assumed that only delocalized imaginary modes are responsible for diffusion. In a similar thought, localized imaginary modes have been instead considered to be resistant to these transport processes (i.e., viscosity) \cite{huang_microscopic_2025}. Based on non-affine viscoelastic response formalism \cite{zaccone_general_2023}, the shear viscosity derived from the imaginary part of the complex shear modulus is given by
\begin{equation}
    \eta = 3\rho \nu(0) \int \frac{g(\omega)\Gamma(\omega)}{m^2\omega^4}d\omega
    \label{Eq: eta_INM}
\end{equation}
where $\rho$ is the number density, $g(\omega)$ is the density of states, $m$ is atomic mass, $\Gamma(\omega)$ is the affine force field correlator, $\nu(0)$ is the zero frequency limit of the spectral density of the memory kernal from the coupling between tagged atoms and thermal environment. Only utilizing imaginary, localized modes (imaginary modes with small participation ratio numbers) into Eq. \ref{Eq: eta_INM}, shear viscosity was found to closely follow temperature trend of independently calculated viscosity from Green-Kubo formalism for metallic liquids and Lennard-Jones liquids \cite{huang_microscopic_2025}. This work similarly demonstrates how unstable modes could be important for transport properties. 

The proposal that diffusion is governed by delocalized imaginary modes while viscosity originates from localized imaginary modes, however, does not fully account for the fact that glass-forming liquids exhibit pronounced dynamic heterogeneity, particularly at temperatures below the Arrhenius–to–non-Arrhenius crossover. Various alternative approaches for relating diffusion coefficients to imaginary modes continue to be proposed \cite{keyes_unstable_2025}. Similarly, another study showed that evaluating viscosity using Eq.~\ref{Eq: eta_INM} while including contributions from all modes—both real and imaginary—yields good agreement with Green–Kubo viscosity in a polymer melt system \cite{singh_viscosity_2025}. Despite these encouraging developments, a microscopic description of viscosity and diffusion in terms of instantaneous normal modes remains elusive as various formulations have been reported to predict against independent molecular dynamics results simultaneously.



In our view, the absence of rigorous connections between instantaneous normal modes and macroscopic material properties stems largely from an incomplete understanding of their fundamental physical nature. To clarify this nature, it is essential to move beyond the predominantly solid-state perspective that has shaped most prior interpretations. In particular, a more balanced framework should also examine the behavior of normal modes in the gas phase, rather than relying on an asymmetric view rooted primarily in crystalline solids. Then, the interpretation of instantaneous normal modes in liquids becomes an interpolation, rather than an extrapolation. 

We recently explored normal mode behaviors in gas systems and their role in inter-particle dynamics \cite{moon_atomic_2023, moon_unifying_2026, moon_heat_2024, moon_normal_2024}. In applying normal modes to dilute gases, it is important to note that the potential cutoff distance should be sufficiently long enough to recover three Goldstone modes. Otherwise, we obtain nearly all modes with zero frequencies. The long cutoff distance rather than a typical few \AA \ short range cutoff distance optimized for solids is physical as real atomic interactions (e.g., wavefunctions) are continuous regardless of how minute they are. 

\begin{figure}
	\centering
	\includegraphics[width=1\linewidth]{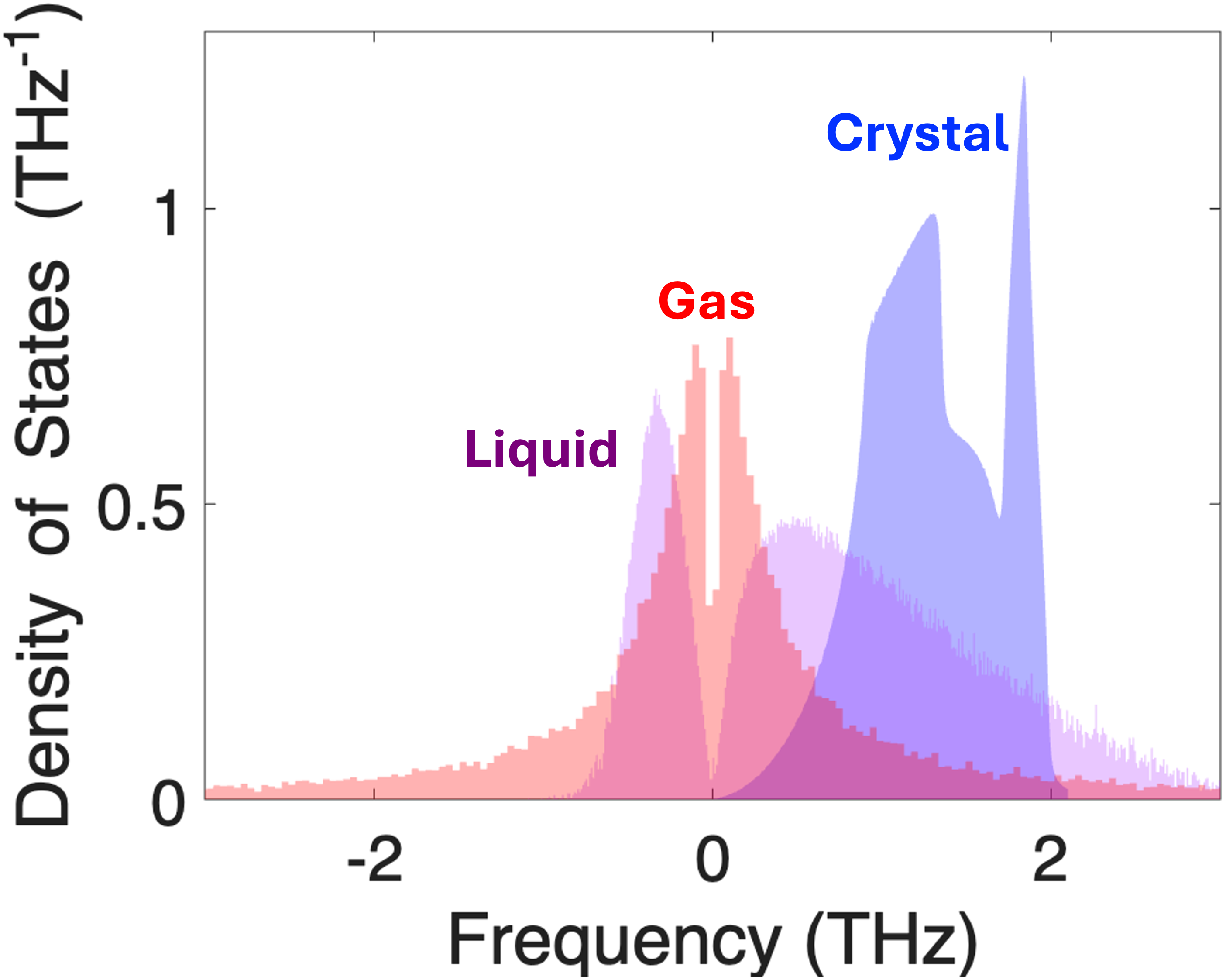}
	\caption{Representative normal mode densities of states for crystal (1 K), liquid (90 K), and gas (120 K) LJ argon at 1 bar. Normal mode
frequencies are multiplied by a factor of 500 for the gas phase for better visualization. This figure is adopted from Ref. \cite{moon_normal_2024}.}
	\label{fig:INM_phase}
\end{figure}

When extrapolating the imaginary mode population dependence on temperature shown in Fig. \ref{fig:Keyes}, it is expected that the imaginary mode population will continue to grow with temperature. Instead, we find that there exists a saturation point. For gas systems without internal degrees of freedom (e.g., intramolecular vibrations) such as LJ argon, we observe approximately the equal number of real and imaginary modes as seen in Fig. \ref{fig:INM_phase}. This observation can be understood as gas atoms having enough kinetic energy to overcome any of the hills and valleys of the potential energy landscape with equal probability \cite{moon_normal_2024}. A majority of frequency magnitudes fall below the GHz range for dilute gases as opposed to the typical THz range for phonons in solids and liquids, due to extremely weak interatomic interactions. Frequency magnitudes become smaller with increase in temperature as interatomic distances become greater \cite{moon_unifying_2026}. Nevertheless, only three Goldstone modes with zero frequency are observed for all temperatures. Further, the observation of significant real mode population in single element dilute gases like LJ argon shown here demonstrates that common assumptions that real modes are vibrational in nature are questionable. 

\begin{figure}[h!]
	\centering
	\includegraphics[width=0.92\linewidth]{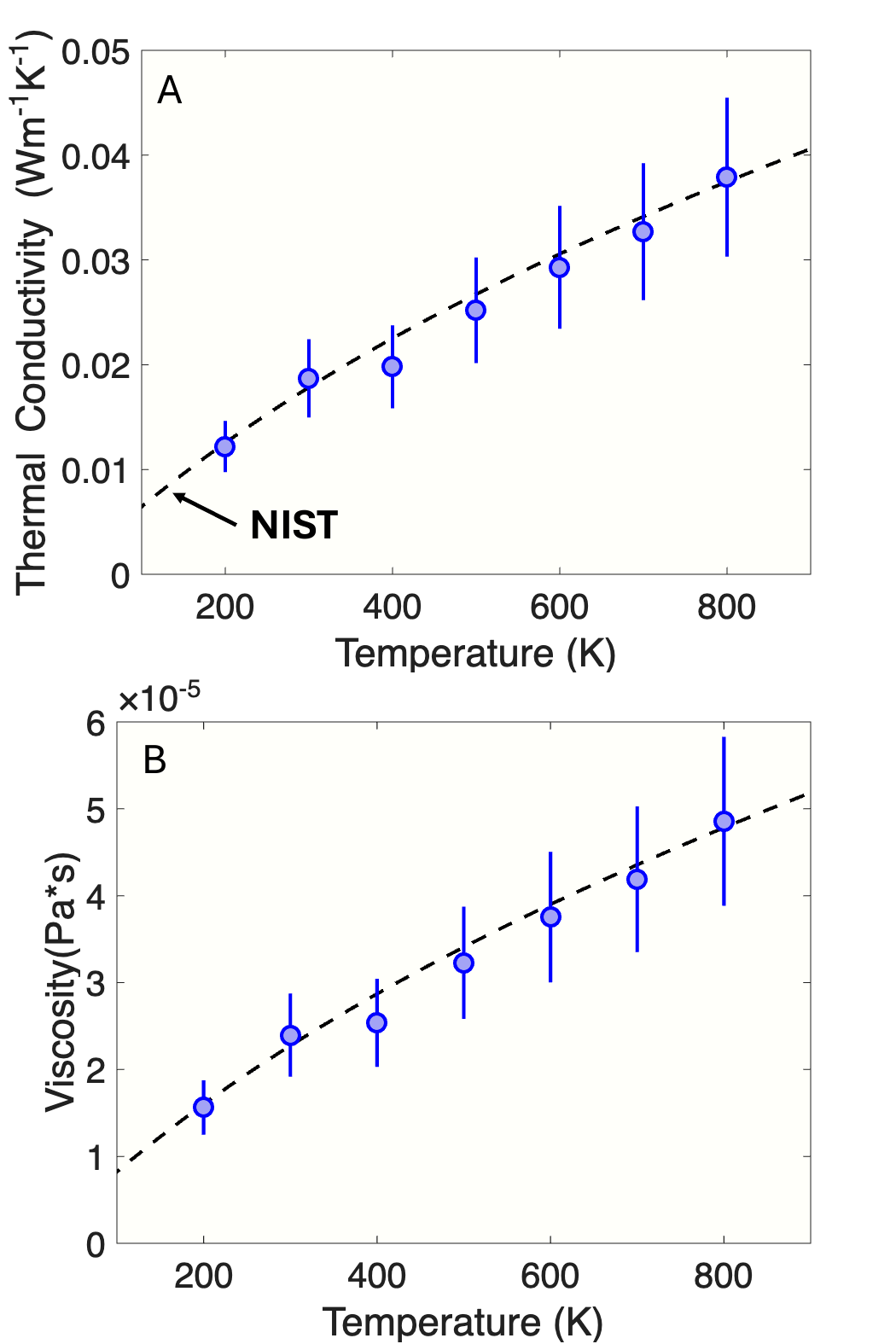}
	\caption{Isobaric thermal conductivity (A) and shear viscosity (B) from 200 K to 800 K at 1 bar for argon gas. Blue circles are from our normal mode calculations and dashed curves are measurement values from NIST \cite{lemmon_thermophysical_2010}. We observe an excellent agreement between our calculations and measurements within a few percent. This figure is from Ref. \cite{moon_unifying_2026}.}
	\label{fig:gas_k_eta}
\end{figure}

To understand the physical nature of instantaneous normal modes in dilute gases, we examined them individually via participation ratios and phase quotients \cite{moon_atomic_2023}. We found that modes begin to depict real-space, local atomic collision and translation motion rather than collective motion of atoms depicting vibrations. To distinguish these modes from phonons, we have named them as collisons and translatons. Further, we showed that eigenvalues in single element dilute gases do not represent square of vibrational frequencies but more generally the curvatures of the potential energy surface \cite{moon_atomic_2023}.

With these new insights about instantaneous normal modes and emergence of gas-like modes, transport properties such as diffusion coefficient, shear viscosity, and thermal conductivity were directly characterized through individual mode kinetic energy lifetimes for both real and imaginary modes as shown in Fig. \ref{fig:gas_k_eta} (diffusion coefficients not shown due to limited available measurements in literature for comparisons). Excellent agreements are demonstrated for all temperatures, demonstrating that reciprocal-space normal modes, often considered as synonyms to phonons, are able to describe atomic dynamics even in dilute gases, merging with real-space kinetic theory of gases.

\begin{figure}
	\centering
	\includegraphics[width=0.90\linewidth]{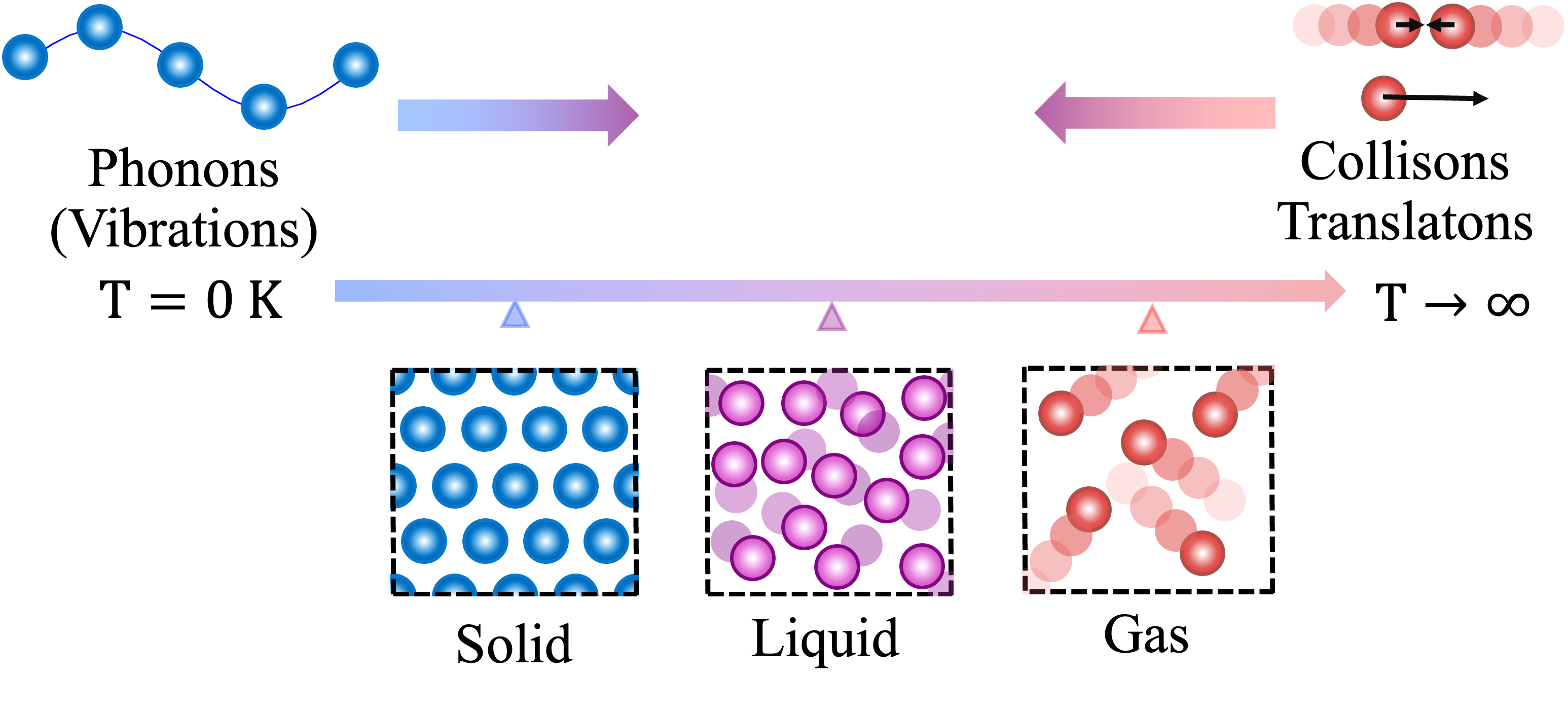}
	\caption{Instantaneous normal mode spectrum across phases from solid to gas. Normal modes in the intermediate such as liquids can perhaps be described as a middle ground between the limits of vibrational modes and gas-like modes.}
	\label{fig:Collison}
\end{figure}

Based on these new findings on the nature of instantaneous normal modes, atomic dynamics in intermediate phases between idealized simple crystalline solids and dilute gases such as liquids can perhaps be described as an interpolation (e.g., hybridized modes) and a generalization of these two limits of normal modes as shown in Fig. \ref{fig:Collison}. 

While normal mode analysis provides a powerful framework for decomposing atomic motion into atomic degrees of freedom, it is not the only approach for characterizing atomic motion in liquids. An alternative and complementary perspective arises from time correlation functions, which describe how microscopic variables evolve over time. Among these, the velocity autocorrelation function plays a particularly important role because it directly connects atomic motion to transport properties such as diffusion. We therefore next examine the velocity autocorrelation spectrum and its interpretation in liquids.

\section{Velocity Autocorrelation Spectra}

In addition to normal modes, velocity autocorrelation spectra, VACF($\omega$), also describe atomic degrees of freedom, as the integral of the spectrum recovers $3N$, where $N$ is the number of atoms. The spectral form of the velocity autocorrelation function contains many features that reflect distinct physical processes. In harmonic solids, it can be shown that VACF($\omega$) is equivalent to the equilibrium normal-mode spectrum \cite{moon_collective_2024, moon_heat_2024}. The VACF formalism is also useful for non-solid systems: integrating VACF($t$) over time yields the diffusion coefficient. Correspondingly, in frequency space through Fourier transform, the zero-frequency limit VACF($\omega = 0$) is directly related to the diffusion coefficient.

\begin{figure}[h!]%
\centering
\includegraphics[width=1\linewidth]{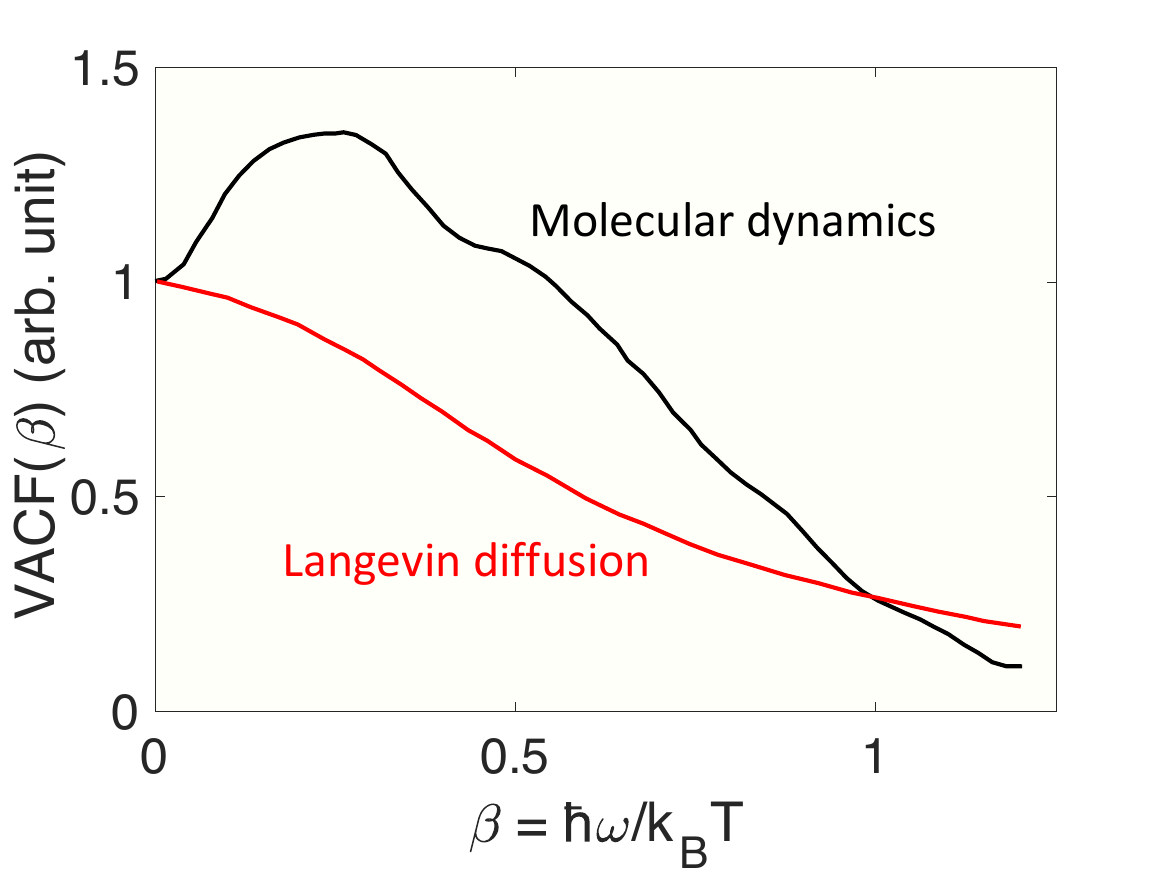}
\caption{Velocity autocorrelation spectra of liquid argon (Lennard-Jones) at 94.4 K, adopted from Ref. \cite{rahman_correlations_1964}. The density is 1.374 g cm\textsuperscript{-3}. Black curve is from molecular dynamics and red curve is a Langevin-type diffusion function of the form, $\frac{\lambda^2}{\lambda^2 + \beta^2}$ where $\lambda$ is related to the diffusion coefficient, $D$ by $\lambda = \hbar/mD$.}
\label{fig:VACF_Rahman}
\end{figure}

These properties have motivated numerous theoretical and experimental efforts to interpret the features of VACF($\omega$) in liquids \cite{rahman_correlations_1964, alder_decay_1970, croxton_introduction_1975, madan_normal_1991, grest_density_1981, williams_velocity_2006, ghosh_molecular_2018, verkerk_velocity_1989}. A seminal example aimed at explaining the overall shape of VACF($\omega$) in liquids was presented by Rahman \cite{rahman_correlations_1964}, illustrated in Fig. \ref{fig:VACF_Rahman}, where VACF spectra obtained from molecular dynamics simulations of a Lennard-Jones argon liquid are compared with predictions from the Langevin diffusion model. The discrepancy between the two curves (notably the peak at $\beta = 0.2$) was attributed to solid-like behavior, namely vibrational motion, persisting in the liquid state. However, the difference at high frequencies (black curve minus red curve) produces unphysically negative values.

\begin{figure}[h!]%
\centering
\includegraphics[width=1\linewidth]{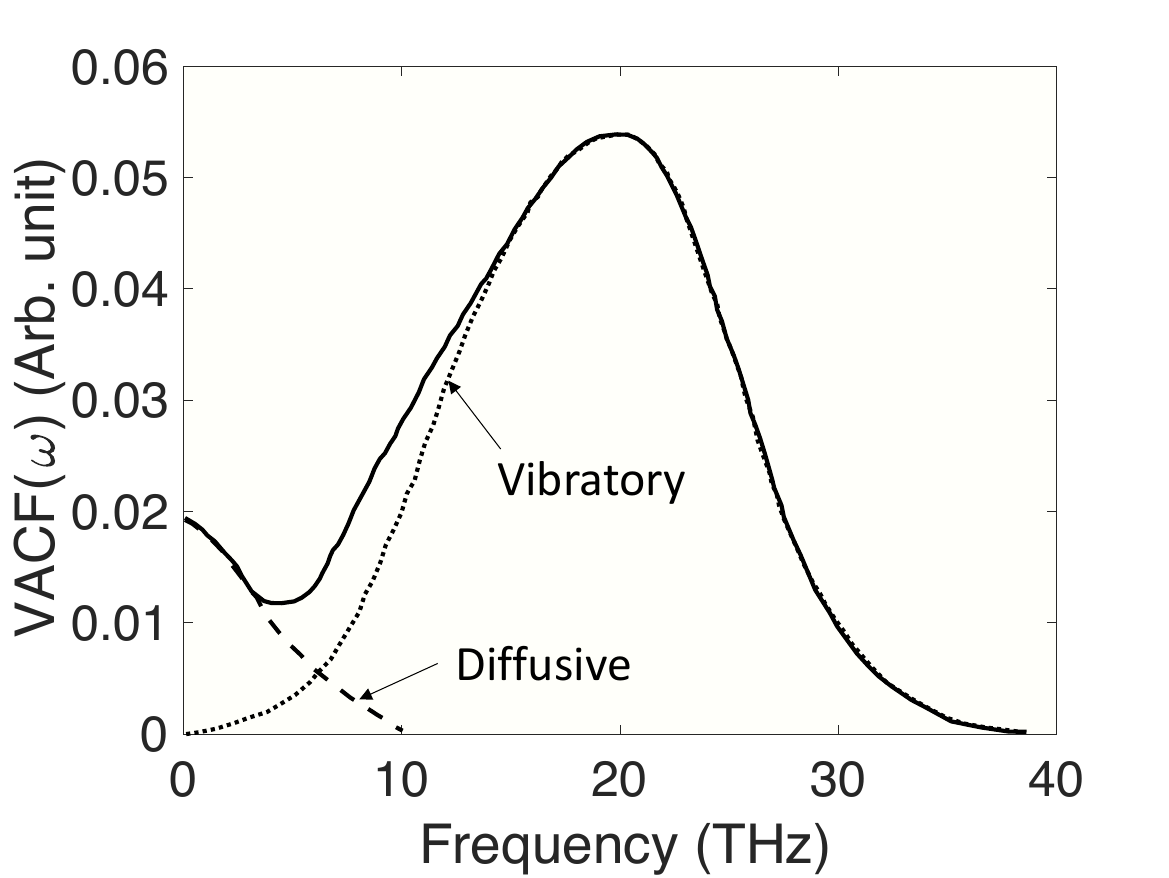}
\caption{Spectral density of liquids sodium at 373 K determined from neutron scattering experiments, adopted from Ref. \cite{croxton_introduction_1975}. Qualitative division of the total spectra into diffusive and vibratory components is proposed.}
\label{fig:VACF_Croxton}
\end{figure}

Motivated by early comparisons between molecular dynamics VACF($\omega$) and theoretical models such as Langevin diffusion and Gaussian spectral distributions, subsequent studies proposed more explicit interpretations of the VACF spectrum. In particular, it was suggested that VACF($\omega$) in liquids can be qualitatively separated into two distinct types of motion: atomic diffusion and vibrations \cite{croxton_introduction_1975}. One such example is illustrated for liquid sodium in Fig. \ref{fig:VACF_Croxton}, where the low-frequency portion of VACF($\omega$) is attributed to diffusive motion, while the medium- to high-frequency region is associated with vibrational motion.

However, this separation was introduced only at a qualitative level, without a quantitative framework to rigorously define the boundary between the two contributions. Based on this interpretation, several liquid-state material properties were proposed to be expressible as the sum of diffusive and vibrational contributions \cite{croxton_introduction_1975}.

Since then, there have been various efforts to quantitatively dissect VACF($\omega$) into these two types motions \cite{lin_two-phase_2003, lin_two-phase_2010, pascal_thermodynamics_2011, madarasz_two_2021, pascal_absolute_2012, huang_absolute_2011}. In particular, Lin et al. evaluated liquid-phase diffusion coefficients by comparing them to the ideal-gas predictions from Chapman–Enskog theory, using this comparison as a quantitative measure of how closely the liquid’s transport behavior approaches that of an ideal gas \cite{lin_two-phase_2003}. As VACF($\omega = 0$) represents the diffusion coefficient, this metric has been used to ascribe diffusion component of the total VACF($\omega$) while the rest of the VACF($\omega$) is considered as vibrations. Using this two-phase model, various thermodynamic properties have been studied. An example is given below for entropy of water \cite{lin_two-phase_2010}. Good agreement between predictions and measurements is demonstrated in a wide range of temperatures. 

\begin{figure}[h!]%
\centering
\includegraphics[width=1\linewidth]{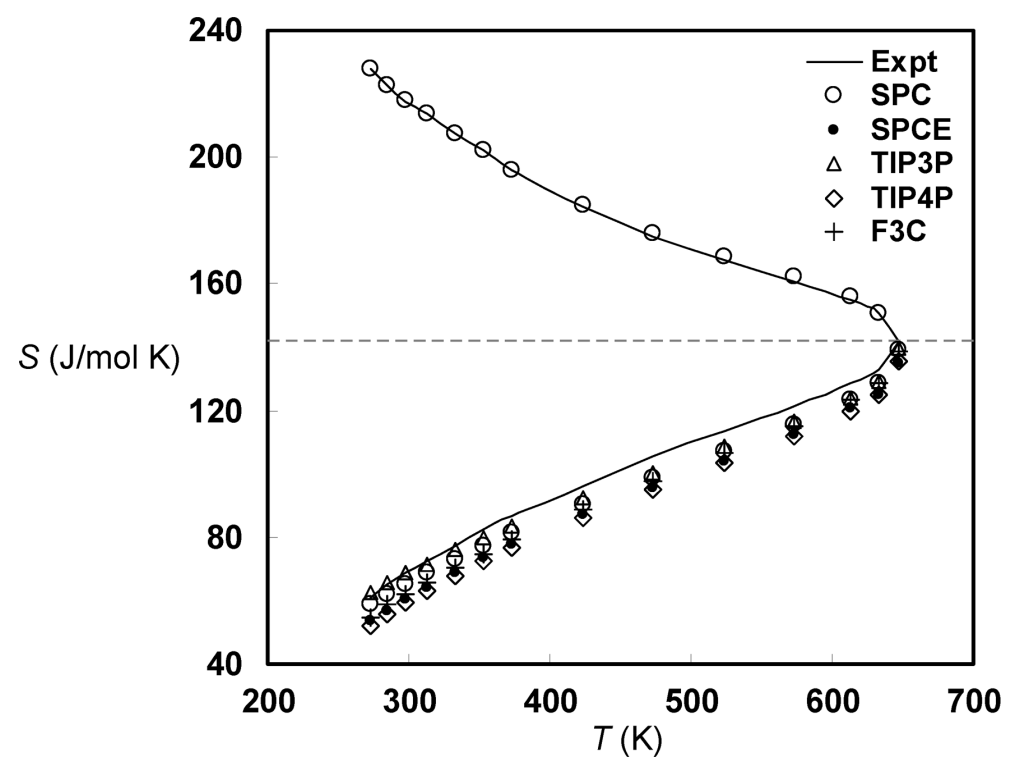}
\caption{Temperature dependent absolute entropy of water. Here, data with markers are based on the two-phase model proposed by Lin et al. \cite{lin_two-phase_2003} with different interatomic potentials and the solid curve is from experiments. Figure is from Ref. \cite{lin_two-phase_2010}.
}
\label{fig:Lin_entropy}
\end{figure}

Despite the success and progress of bisecting the VACF($\omega$) into the two components in describing properties of liquids, various uncertainties and problems remain. Some examples include: VACF($\omega$) of hard sphere gases can decline more slowly in frequency than that of the actual liquid, causing some negative VACF($\omega$) for solids \cite{desjarlais_first-principles_2013}. Real gases with finite interatomic interactions do not behave as hard sphere gases and anharmonicity in vibrations including frequency shifts are not naturally included in the subtraction. Finally, microscopic level of understanding of liquid properties is limited as VACF($\omega$) is described only in terms of two diffusive and vibrational components rather than individual degrees of freedom. 

Although quantities such as the velocity autocorrelation function provide valuable information about atomic dynamics in simulations, experimental validation requires techniques capable of probing microscopic motion directly. Scattering methods using X-rays and neutrons offer precisely this capability by measuring the dynamic structure factor, which encodes atomic correlations in both space and time. In the next section, we discuss how these experimental techniques provide insight into liquid dynamics.

\section{X-ray and Neutron Scattering}

\begin{figure}
\centering
\includegraphics[width=1\linewidth]{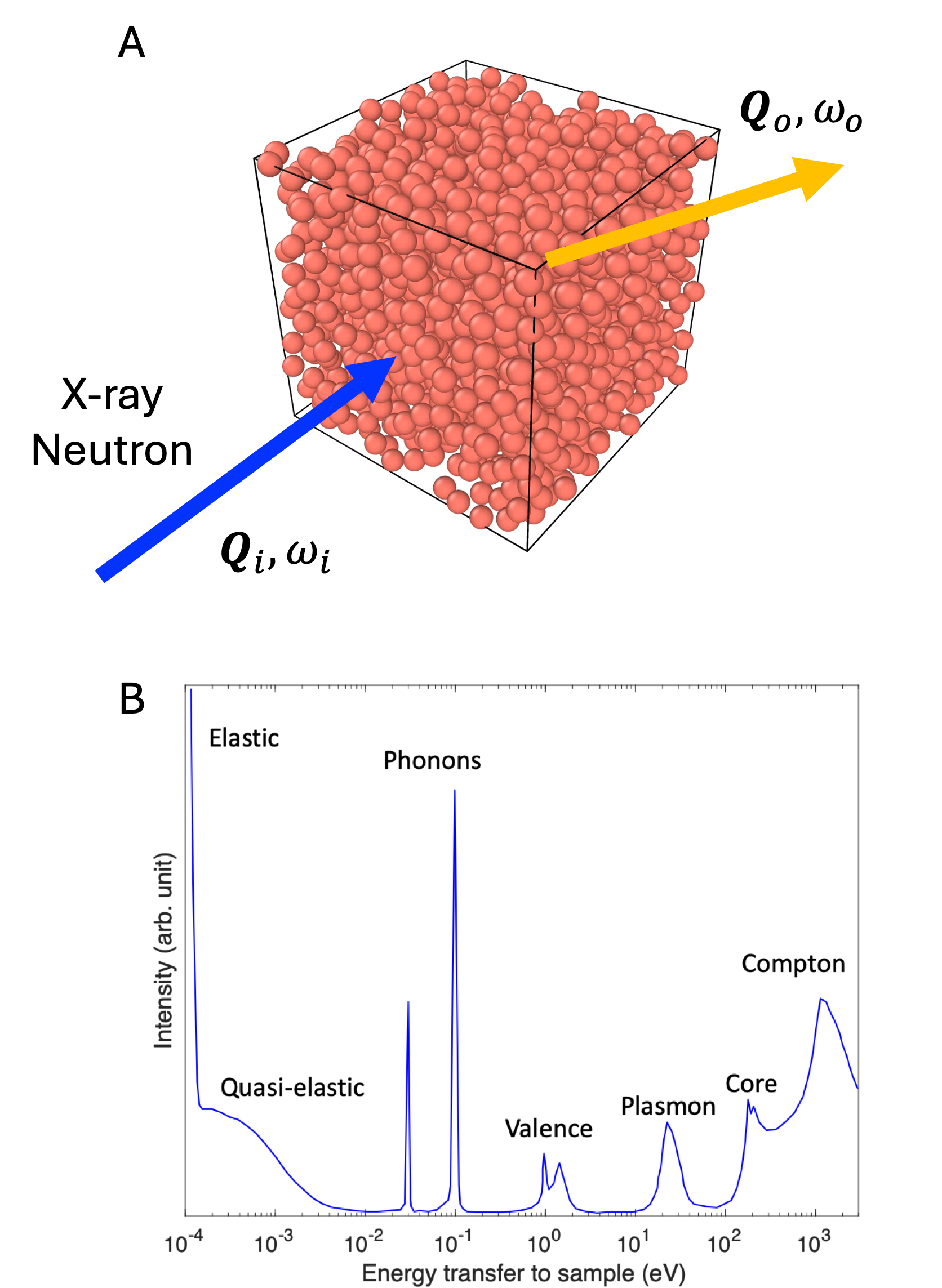}
\caption{(A) Schematic of X-ray and neutron scattering to probe atomic dynamics in material of interest. (B) Various excitations that can be studied through scattering measurements. (B) is adopted from Ref. \cite{baron_high-resolution_2015}.
}
\label{fig:Inelastic}
\end{figure}

To study dynamics at the atomic scale experimentally, X-rays and neutrons are often used as probes due to their short wavelength $\lesssim$ 1 nm. We focus our discussion on understanding atomic dynamics in liquids through X-ray and neutron scattering \cite{sette_collective_1995,cowley_inelastic_1971, brockhouse_time-dependent_1959, cunsolo_terahertz_2017, iwashita_seeing_2017, ruiz-martin_high-frequency_2006, scopigno_density_2000, sette_dynamics_1998, scopigno_high-frequency_2002, carneiro_velocity-autocorrelation_1976, cunsolo_transverse_2012, trigo_fourier-transform_2013}. A simple working principle of X-ray and neutron scattering is demonstrated in Fig. \ref{fig:Inelastic}A. Here, X-rays or neutrons with known wavevectors ($\boldsymbol{Q}_i$) and frequency ($\omega_i$) or energy are incident on a material of interest. After interacting with the material, X-rays and neutrons are collected by a detector and their wavevector ($\boldsymbol{Q}_o$) and frequency ($\omega_o$) are characterized. Changes in the wavevector ($\boldsymbol{Q}=\boldsymbol{Q}_o - \boldsymbol{Q}_i$), frequency ($\omega = \omega_o-\omega_i$), and intensities due to matter interaction are encoded in a quantity known as dynamic structure factor (DSF, $S(\boldsymbol{Q}, \omega)$) and provide information regarding excitations within the material. Types of excitations that can be measured by scattering vary significantly as shown in Fig. \ref{fig:Inelastic}B \cite{baron_high-resolution_2015}. For studying atomic dynamics in liquids, probing up to $<$ $O$(100) meV is typically sufficient.

Dynamic structure factor is defined as
\begin{equation}
    S(\boldsymbol{Q}, \omega) = \frac{1}{2\pi N} \sum_{i,j}\int \big\langle e^{i\boldsymbol{Q}\cdot [\boldsymbol{r}_i(0)-\boldsymbol{r}_j(t)]} \big\rangle e^{i\omega t}dt 
    \label{Eq: DSF}
\end{equation}
where $\boldsymbol{r}_i$ is the position of the $i$th atom at time $t$ and angled brackets mean thermal and ensemble average. Since time dependent atomic trajectories are provided in molecular dynamics, dynamic structure factor can also be calculated \cite{moon_propagating_2018, moon_thermal_2019, bryk_ab_2017, bryk_collective_2014, beltukov_boson_2016}. In crystalline solids, the wavevector- and frequency-resolved intensity peaks of the dynamic structure factor depict the phonon dispersion relations \cite{ashcroft_solid_1976, krisch_inelastic_2006, sinha_theory_2001, burkel_phonon_2000}. The peak position gives the phonon frequency at a given wavevector, the peak width (full width at half maximum) is inversely proportional to the phonon lifetime, and the slope of the dispersion curve corresponds to the phonon group velocity. 

\begin{figure}
\centering
\includegraphics[width=1\linewidth]{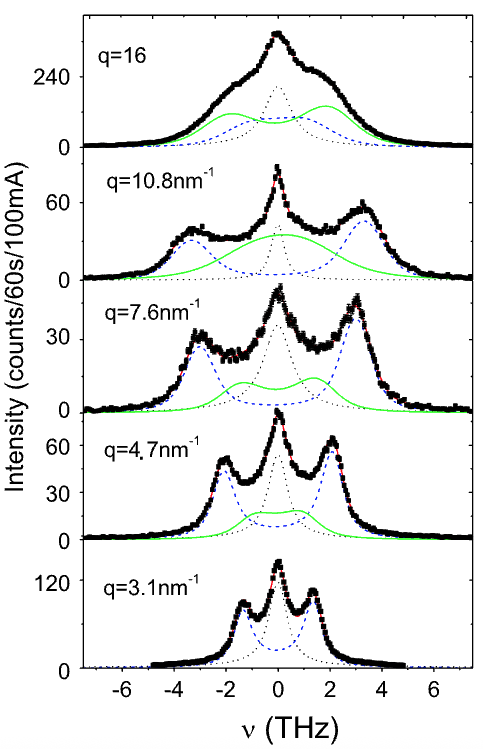}
\caption{Inelastic X-ray scattering spectra of sodium at 393 K in the liquid phase at the indicated wavevector values. The quasi-elastic (dotted black line) and inelastic transverse(dot-dashed green line) and longitudinal (dashed blue line)components of the fitting model are also reported afterconvolution to the instrumental function. The figure is adopted from Ref. \cite{giordano_fingerprints_2010}.
}
\label{fig:Monaco}
\end{figure}

As liquids are usually isotropic, orientational averaging simplifies $S(\boldsymbol{Q}, \omega)$ to $S(Q,\omega)$. Similar to crystalline solids, $S(Q,\omega)$ provides information regarding phonon-like vibrations such as acoustic excitations as shown in Fig. \ref{fig:Monaco}.  We observe propagation of longitudinal and transverse acoustic excitations up to $\sim$ 8 nm\textsuperscript{-1} and 10 nm\textsuperscript{-1}, respectively. Above these wavevectors, dispersion appears to fold  \cite{giordano_fingerprints_2010}. There is no clear Brillouin zone unlike crystalline solids but rather a pseudo-Brillouin zone set by the first diffraction peak ($\sim$ 20.2 nm\textsuperscript{-1} for liquid sodium shown here) \cite{giordano_universal_2009}. Dispersion folding occurs approximately at half the pseudo-Brillouin zone.

In addition to characterizing acoustic excitations themselves, dynamic structure factor also provides important information about liquid viscoelasticity, including positive sound dispersion \cite{majumdar_two-component_2026, ruocco_history_2008, cunsolo_microscopic_2001} and emergence of a momentum gap \cite{khusnutdinoff_collective_2020, yang_emergence_2017-1, baggioli_gapped_2020, baggioli_how_2021, egelstaff_introduction_1994}. In many liquids, both experiments and simulations have reported an increase in longitudinal sound velocity with increasing wavevector relative to its macroscopic adiabatic value. This enhancement is  attributed to the coupling between structural relaxation processes and the density fluctuations associated with propagating acoustic waves \cite{cunsolo_microscopic_2001}. This transition from hydrodynamic sound to "fast" sound is considered as changes from solid-like to liquid-like atomic dynamics. 

\begin{figure}
\centering
\includegraphics[width=1\linewidth]{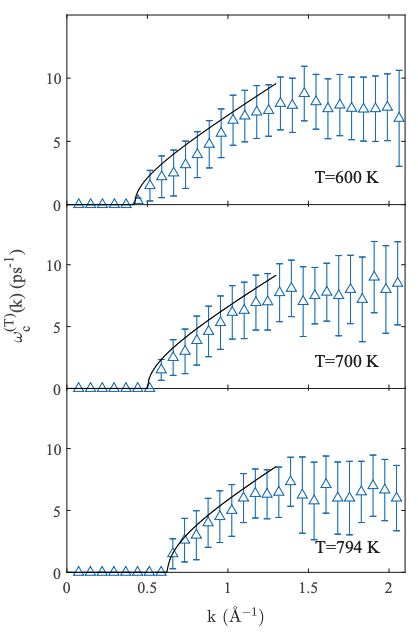}
\caption{Transverse acoustic excitation dispersion of liquid gallium at various temperatures: Open triangles show the results of molecular dynamics simulation and solid curves represent theoretical
results. Gap in the wavevector (momentum gap) is demonstrated at zero frequency. The figure is adopted from Ref. \cite{khusnutdinoff_collective_2020}.
}
\label{fig:k_gap}
\end{figure}

Similar arguments are made to transverse excitations. Using Maxwell interpretation to describe the viscoelastic response in liquids, we expect a gap ($k_{gap}$) in the transverse acoustic excitation dispersion at zero frequency \cite{egelstaff_introduction_1994}. This is demonstrated in independent molecular dynamics calculations as shown in Fig. \ref{fig:k_gap} for liquid gallium \cite{khusnutdinoff_collective_2020}. The momentum gap scales inversely with Maxwell relaxation time characterizing shear viscosity; therefore, we expect the gap widening with increase in temperature.

As demonstrated by these examples, various atomic dynamics in liquids including structural relaxation and vibrations is represented in the dynamic structure factor in the reciprocal space. Depending on the availability of frequency or wavevector ranges in the data, it often helps to interpret dynamic structure factor in real-space and/or time domains through Fourier transform from a different viewing angle. Dynamic pair density function can be obtained by Fourier transforming the dynamic structure factor in space by
\begin{equation}
    g(r,\omega)=\frac{1}{2\pi^2nr}\int S(Q,\omega)Q\sin(Qr)dQ
\end{equation}
where $n$ is the atomic number density. Pair distribution function, $g(r)$, is obtained by integrating $g(r,E)$ by frequency and static structure factor, $S(Q)$ is, in turn, Fourier transform of $g(r)$ in space \cite{egami_dynamic_2012}. 

\begin{figure}
\centering
\includegraphics[width=1\linewidth]{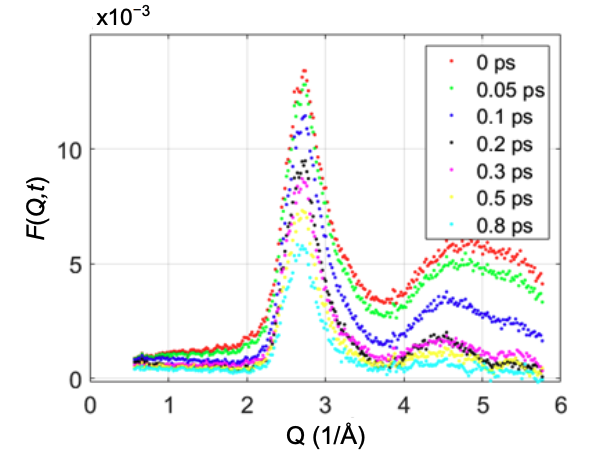}
\caption{Intermediate scattering function for Cu\textsubscript{49}Zr\textsubscript{45}Al\textsubscript{6} liquid at 1060 $^{\circ}$C, measured by inelastic neutron scattering.  The figure is adopted from Ref. \cite{kalicki_correlated_2025}.
}
\label{fig:ISF}
\end{figure}

When Fourier transforming the dynamic structure factor in time instead, Intermediate scattering function (ISF) is obtained as
\begin{equation}
    F(Q,t) = \int S(Q,\omega) e^{i\omega t}d\omega.
\end{equation}
Here, $F(Q,t=0)$ is equivalent to $S(Q)$. $F(Q,t)$ has two parts: self-part and distinct part when $i=j$ and $i \neq j$ in Eq. \ref{Eq: DSF}, respectively. An example of ISF measurements by inelastic neutron scattering is demonstrated in Fig. \ref{fig:ISF} for a metallic liquid \cite{kalicki_correlated_2025}. Intermediate scattering function can also be measured by neutron spin-echo \cite{mezei_principles_1980} and X-ray photon correlation spectroscopy \cite{frey_liquid-like_2025}. Different rates of intensity decays are observed depending on the wavevector. In particular, the first sharp diffraction peak (FSDP) has gained a lot of interests for the last few decades, which is widely considered to represent medium range structural order \cite{wilson_prepeaks_1994, ryu_curie-weiss_2019, dzugutov_medium-range_1998, elliott_origin_1992, salmon_real_1994}. For the metallic liquid shown here, FSDP occurs at $Q = Q_1$ around 2.7 \AA\textsuperscript{-1}. Intensity of $F(Q=Q_1,t)$ usually decays with time in two steps: Fast relaxation describing vibrations and short range $\beta$-relaxation followed by a slower relaxation describing $\alpha$-relaxation \cite{ryu_alpha-relaxation_2025}. $\alpha$-relaxation is directly related to shear viscosity and its relaxation time, $\tau_\alpha$, is generally regarded as the relaxation time of the atomic cage surrounding an atom \cite{jaiswal_atomic-scale_2015}. $\beta$-relaxation is often considered as the precursor to $\alpha$-relaxation \cite{yu__2013}. ISF, particularly at $Q_1$, has been widely used to study glass transition and Arrhenius to super-Arrhenius viscosity changes in liquids \cite{scalliet_thirty_2022, jaiswal_atomic-scale_2015, ryu_alpha-relaxation_2025, wang_dynamic_2019, puosi_dynamical_2018, gnan_microscopic_2019}. 

When Fourier transforming the dynamic structure factor in both time and space, we obtain Van Hove function (VHF) \cite{van_hove_correlations_1954} as
\begin{equation}
\begin{split}
    G(r,t)&= 1 + \frac{1}{2\pi^2nr} \int S(Q,\omega) Q\sin(Qr)e^{i\omega t} dQd\omega \\
          &=\frac{1}{4\pi nNr^2} \sum_{i,j} \langle \delta(r-|\boldsymbol{r}_i(0) - \boldsymbol{r}_j(t)| \rangle.
    \end{split}
\end{equation}
At $t=0$, VHF is the pair distribution function: $G(r,0) = g(r)$ and the VHF at other times describes how the pair distribution evolves with time. The VHF can also be separated into two terms, the self-part ($i=j$) and the distinct part ($i\neq j)$ as
\begin{equation}
    G(r,t) = G_s(r,t) + G_d(r,t)
\end{equation}
where the self-part, $G_s(r,t)$ describes how an atom moves away from its initial position at $t=0$ while the distinct part, $G_d(r,t)$, depicts how an atom $j$ moves from its initial position measured from the initial position of atom $i$. 

While the VHF has been measured decades ago \cite{brockhouse_time-dependent_1959}, it was only recent that measurements over a wide range of $Q$ and $\omega$ became feasible within a beamtime of a few days \cite{iwashita_seeing_2017, ashcraft_experimental_2020,shinohara_real-space_2022,shinohara_dynamical_2025}. Recent demonstrations of the experimental determination of VHF by inelastic X-ray and neutron scattering have revived the interest in the VHF, particularly for distinct parts as the VHF describes atomic dynamics in liquids in real space and time that can often be more intuitive to interpret than dynamic structure factors \cite{iwashita_seeing_2017}.

\begin{figure*}
	\centering
	\includegraphics[width=0.85\textwidth]{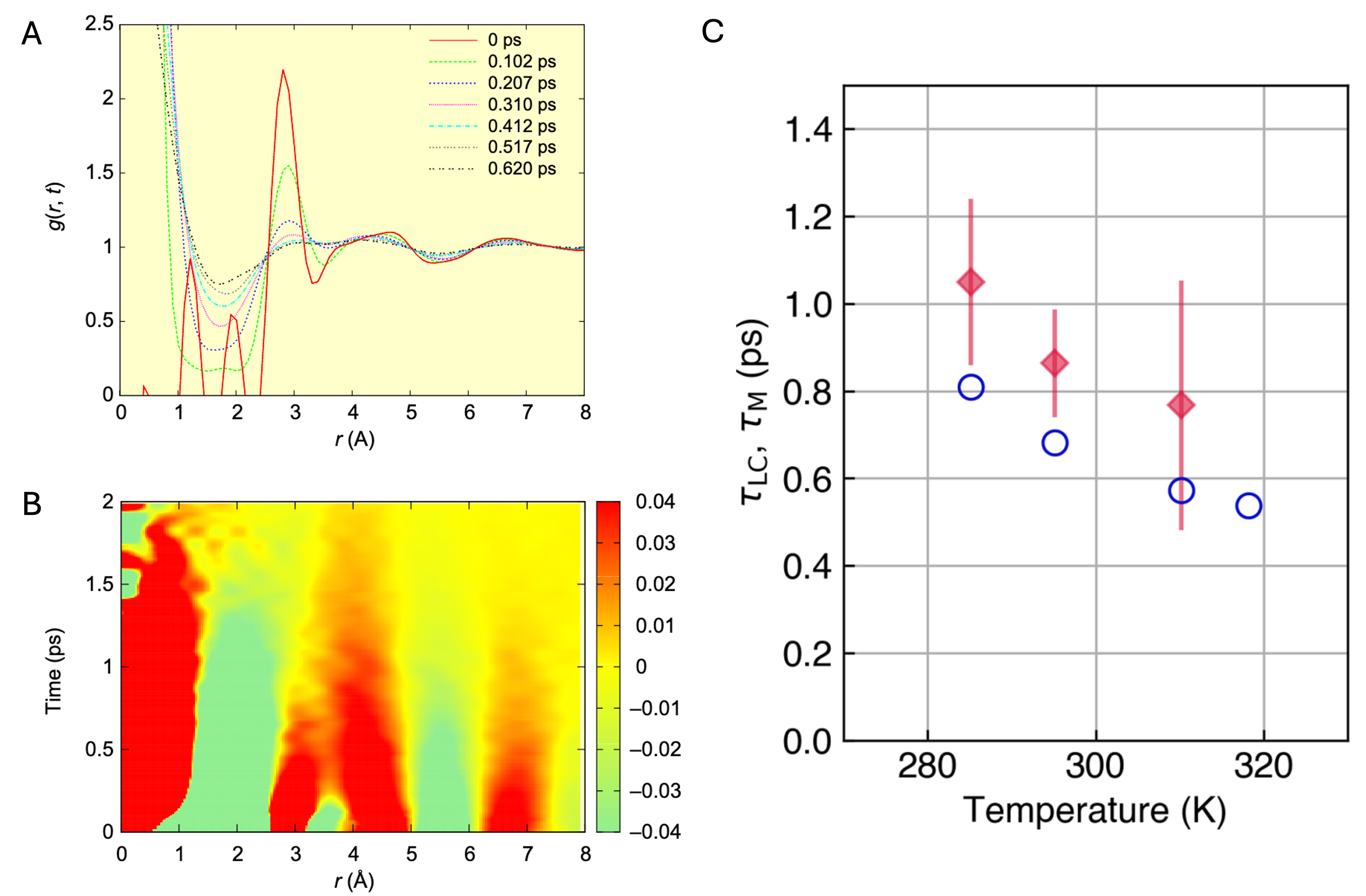}
	\caption{Van Hove Function (VHF) results for water determined by inelastic X-ray scattering. The notation of $g(r,t)$ from Ref. \cite{iwashita_seeing_2017} is the same as $G(r,t)$ in this Review. (A) VHF at different times, (B) 2D plot of $G(r,t)-1$, and (C) temperature dependent Maxwell relaxation time ($\tau_M$) characterizing shear viscosity and average time to loose a neighboring atom inferred from VHF ($\tau_{LC}$). The figure is adopted from Ref. \cite{iwashita_seeing_2017} and Ref. \cite{shinohara_viscosity_2018}}
	\label{fig:VHF}
\end{figure*}

 The VHF results for water as an example obtained by inelastic X-ray scattering are shown in Fig. \ref{fig:VHF} \cite{iwashita_seeing_2017, shinohara_viscosity_2018}. The VHF obtained here mainly describes the correlation between oxygen ions due to small cross-section of hydrogen to X-rays. Self-part of the VHF dominates the VHF up to 1.5 \AA \ with sharp peak around $r = 0$ depicting self-diffusion and the distinct part beyond 1.5 \AA \ displays collective motion. As clearly seen in Fig. \ref{fig:VHF}B, the first peak at 2.8 \AA \ at $t=0$ merges with the second peak located between 4 \AA \ and 5 \AA \ with increase in time. This behavior in in strong contrast with metallic liquids where the VHF decays monotonically and the peak positions remain distinct. This difference has been attributed to stark difference in local atomic environment: Atoms in metallic liquids are surrounded by approximately 12 other atoms as nearest neighbors whereas the coordination number is around 4 for liquid water \cite{iwashita_seeing_2017}. Thus, a more spatially correlated rearrangement is needed for water as a molecule loses its neighbor.

In addition to the time dependent VHF peak locations, relaxation times of the VHF peaks and valleys (area or peak intensity) provide a direct glimpse at the atomic dynamics in liquids. Prior works have demonstrated that these relaxation times are proportional to $\tau_{LC}$ representing an average time for atoms to lose a neighboring atom \cite{egami_correlated_2020}. For water, a second slowly decaying relaxation time of the first peak of the VHF has been shown to be nearly equivalent to $\tau_{LC}$, which in turn is approximately the Maxwell relaxation time characterizing shear viscosity as shown in Fig. \ref{fig:VHF} C \cite{shinohara_viscosity_2018}. Similar to these water results, VHF has been utilized both experimentally and computationally to understand atomic dynamics in other liquids such as aqueous liquids, metallic liquids, molten salts, and others in real space and time \cite{shinohara_real-space_2022, shinohara_identifying_2019}.

In principle, if a wide enough range of $Q$ and $\omega$ is probed, dynamic structure factor and Van Hove function have the same information but they each provide a different angle of perspectives. As the examples given in this Review demonstrates, the dynamic structure factor may be a natural choice of analysis probing atomic vibrations as it is in the reciprocal space whereas the Van Hove function could provide a more intuitive picture of atomic interactions and structural relaxations in liquids. 

As illustrated in the examples above, modern scattering experiments provide increasingly detailed views of atomic dynamics in liquids, complementing insights obtained from theoretical analysis and molecular simulations. Together, these approaches are beginning to reveal connections between microscopic dynamics and macroscopic material properties. In the final section, we briefly discuss emerging opportunities and future directions for advancing the microscopic understanding of liquids.

\section{Forward}

Liquids occupy a unique position among the phases of matter. They are dynamically disordered like gases yet remain densely packed and strongly interacting like solids. Because of this dual character, liquids resist simple theoretical descriptions that make solids and gases tractable.

Historically, the physics of liquids has often been approached by extending ideas developed for other phases of matter. Some frameworks treated atomic motion in liquids as an extension of gas dynamics (e.g., ideal gas law), while others have approached from solid perspectives (e.g., lattice theories). These perspectives have each provided valuable insights, yet neither alone fully captures the nature of liquid dynamics. 

Modern theoretical, computational, and experimental approaches increasingly suggest that liquids appear to exist in a regime where these two pictures merge, and where the distinction between vibrational and diffusive motion becomes blurred. The dynamics of liquids may be understood in terms of hybrid excitations that interpolate between solid-like and gas-like behavior. From this perspective, the traditional distinction between vibrational and collisional motions may not represent two fundamentally different processes, but rather different manifestations of the same underlying dynamical landscape. Developing frameworks capable of describing this continuity of motion remains one of the central challenges in liquid state physics.

As this Review has shown, a substantial gap still exists between theoretical and computational descriptions of liquids and their experimental characterization. In crystalline solids, clear correspondences exist between physical concepts (vibrations as heat carriers), theoretical constructs (normal modes), and experimental measurements (dispersion relations). For liquids, such direct connections are less well defined. Nevertheless, recent advances are rapidly narrowing this divide. High-performance computing now enables atomistic simulations that approach experimental length and time scales, while modern scattering techniques probe atomic motion with unprecedented spatial and temporal resolution. Together, these developments offer the possibility of directly linking theoretical descriptions of microscopic motion with experimentally observable quantities. 

Ultimately, the study of liquids touches on broader questions about how to describe atomic motion that governs thermodynamics and materials properties in the most general way.  Although significant challenges remain, recent advances in theory, computations, and experiments described in this Review suggest we are getting close.

\bibliography{CPR.bib}

\end{document}